# Documenting Spreadsheets

Raymond Payette, MPA,
3475 Vautelet #306,
Quebec, Que. CANADA
G1W 4V9
Support@excelhawk.com
© Raymond Payette 2006

**ABSTRACT**

This paper discusses spreadsheets documentation and new means to achieve this end by using Excel's built-in "Comment" function. By structuring comments, they can be used as an essential tool to fully explain spreadsheet. This will greatly facilitate spreadsheet change control, risk management and auditing. It will fill a crucial gap in corporate governance by adding essential information that can be managed in order to satisfy internal controls and accountability standards.

**INTRODUCTION**

The necessity of spreadsheet documentation has made by institutions responsible for establishing standards.  Recognized authorities have expounded the documentation's content. Documentation methods cover external file and internal file documenting, the subjective evaluation, internal data documentation and macro documentation.

**1. THE NECESSITY**

Information and Communication is an essential principle of the Internal framework of the Internal Control proposed by COSO. Similarly communication is one of the five fundamental components of the COBIT Framework issued by the IT Governance Institute and documentation is one of the fundamental means of communication.
Policy and procedures should be documented, so should the spreadsheets.

The documentation objective is to ensure compliance to corporate governance. It is required to prove the basis of corporate decisions.

**2. DOCUMENTATION CONTENT**

According to *Guide for Preparing, Documenting and Referencing,* the goal is to facilitate quality assurance by the use of checklists, crossed-indexed tables of content and other processes.

The content should specify:
"Job title and code,
Title of the Spreadsheet
Work-paper index, specifying the data's location, the name of the person updating the spreadsheet with its date
Its purpose
The preparer, his title and the date
The reviewer's name and the date

The data's source
Data verification such as sampling
Spreadsheet layout
Printing…"
This documentation can be internal to the spreadsheet or external.

According to *Spreadsheet Modeling for Best Practice*, good documentation
"Allows another person to use a model and what it is doing even if the model has been out of use for some time;
reduces the risk that the model falls into disuse because only one or two people know how to use or trust its results; and
reduces the number of irritating interruptions for the model developer, after handing over the responsibility for the model."

According to *Methodology for the Audit of Spreadsheet Models*, by H.M. Customs & Excise the developer should make it clear:
"what it's for;
what it does;
how it does it;
what assumptions were made in its design;
what constants are used and where they are held;
who developed it;
when;
when and how it has been changed since being brought into use;
the presence and purpose of any macros."

## 3. DOCUMENTATION METHODS

### 3.1 External file documentation

Usually « versioning » is done this way by keeping file inventory and recording any change made to it. This entails checkout and check-in controls and may involve file sharing and conflict resolution. Microsoft has addressed this problem as described in *Sarbanes-Oxley: Addressing Sarbanes-Oxley Challenges Using the Microsoft Office System*. It recommends the use of SharePoint® and InfoPath® that imply using Windows Server 2003® and SQL Server 2000®.

IBM offers onsite or offsite data retention with Tivoli server. There are many companies that offer backup services that can be used for versioning. An extranet can also be used for this purpose. Though files are backed up there seems to be no documentation on the files themselves so there remains a need to explain the content of spreadsheet files.

A file inventory could be made and commented every time there are changes. For example the following information could be recorded:
Path & name:
Date and Time stamp:
Attribute: (read-Write/Read-only/Archive/Hide)
Size:

The following information could be added manually:
Author:
Purpose:
Type: (Current/Active/Standby/Archive/Backup)
Security: (Secret/Confidential/Private/Colleagues/Entity/Public)

Access: (Unique/Restricted/Unrestricted) --- Information Rights Management
Reason of change: (Creation/Modification/Update/Addition/Deletion)

**3.2 Subjective information**

One of the main problems is to obtain information from data. The best way to do so is simply to have the author add the subjective information. There are some cases where the attempted goal has not been reached and so the author's intent becomes invaluable to the spreadsheet's appraisal. Producing significant documentation solely by computational means is neither realistic nor efficient. Complete documentation requires human intervention. This should be made as easy as possible and should limit itself to useful information. Too much data and information may be counter-productive.

**3.3 Internal file documentation**

Excel can be used to manage the inventory of Excel files. For example the software SCANXLS produces the following information for each Excel File:

| | |
|---|---|
| FullPath | Full path and file name |
| Filename | File name only. Convenient to sort on this column to find duplicates. |
| ScanTime | This might be useful to estimate completion times for other scans. |
| Size | Filesize in bytes |
| Created | Date created |
| Accessed | Date last accessed |
| Modified | Date last modified |
| Attributes | A=Archive, H=Hidden, R=Read-only, S=System |
| Observations | Any errors or other problems opening or closing the workbook or its referenced workbooks. |
| FileFormat | The format of the workbook, usually the version of Microsoft Excel it was saved as. |
| Contents | The value of an expression in the first active worksheet, for identification purposes, default is A1 |
| Author | File Property Author |
| Manager | File Property Manager |
| No. Links | Count of links to other workbooks |
| Linked files | Full path and file name of linked workbooks |
| NeedRecalc | Needs a recalculation |
| Backup | Whether it creates a backup when saved (recommended) |
| NrUWB | Count of workbooks with "unusual" settings; see next for the list of settings reported. |
| UnusualWB | "Accepts labels in formulas", " Custom Document Properties ", "Excel 4 Macro sheets", "Has Routing Slip ", "Is running as an Add-In", "Multi User Editing", "Precision As Displayed "Remove Personal Information is enabled", "VBA code has been digitally Signed " |
| NrUWS | Count of worksheets with "unusual" settings; see next for the list of settings reported. |
| UnusualWS | "Circular Reference", "Consolidation sources", "Filtered with hidden rows", "OLE Objects", "Pivot Tables", "Protected", "Query tables", "Scenarios", "Lotus evaluation rules", "Lotus formula entry", "Hidden / Very Hidden" |
| Names | Count of range names in the workbook |
| Worksheets | Count of worksheets |
| Find1.4 | Count of cells found containing the text specified in green-shaded cell in row 1 of each "Find" column |
| Where1.4 | Location of a cell with the longest formula with that found text. A rough idea of the complexity of the model logic. |
| Code | Total count of lines of VBA Code in all modules |
| Validation | Count of cells with validation criteria |
| Comments | Count of cells with comments |
| Constants | Count of cells with constant (literal) values |
| Numbers | Count of numeric input cells |
| Formulas | Count of formulas |
| Errors | Count of formulas evaluating to #error values |
| ErrorVal | Count of cells evaluating to error. This may sometimes differ from that above where there are multiple failures of error checking. |
| TextDate | Count of cells with text contents that resemble a date |
| TextNum | Count of cells with text contents that resemble a number |
| Inconsis | Count of formulas inconsistent with the region they are in |
| OmitsCells | Count of formulas that omit nearby cells. |
| ULformula | Count of unlocked formulas |
| EmptyRef | Count of formulas referencing empty cells |
| Scoring | This is an arbitrary function that adds up the risk factors in the preceding columns. |

The advantage of this form of versioning is to evaluate the spreadsheet's risk. Since all results are factual, they do not contain subjective opinions such as the purpose of the spreadsheet that could be added and could significantly improve its usefulness.

**3.4 Internal data documentation**

Excel provides a way to track the spreadsheet's internal history, with the **Tools**, **Track Changes** and **Highlight Changes** menu. The file must have previously been identified as a shared workbook.

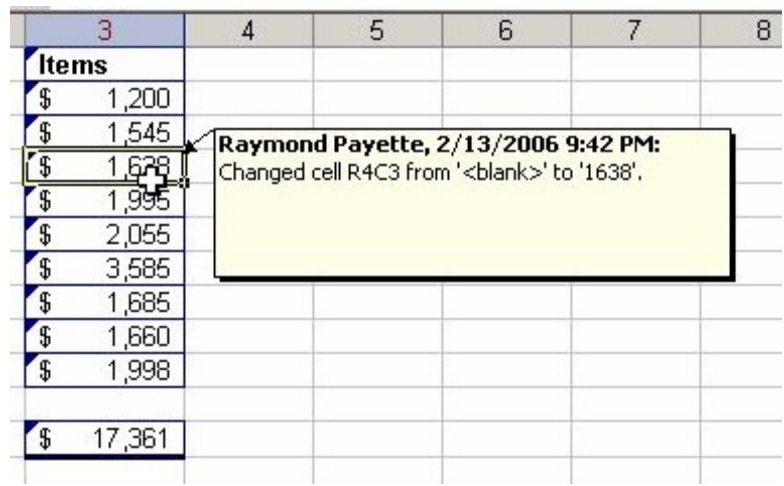

We can choose to List the changes on a new sheet that will produce:

| Action Number | Date | Time | Who | Change | Sheet | Range | New Value | Old Value | Action Type | Losing Action |
|---|---|---|---|---|---|---|---|---|---|---|
| 1 | 2/13/2006 | 9:45 PM | Raymond Payette | Cell Change | Sheet1 | R1C3 | Items | <blank> | | |
| 2 | 2/13/2006 | 9:45 PM | Raymond Payette | Row Auto-Insert | Sheet1 | R2 | | | | |
| 3 | 2/13/2006 | 9:45 PM | Raymond Payette | Row Auto-Insert | Sheet1 | R3 | | | | |
| 4 | 2/13/2006 | 9:45 PM | Raymond Payette | Cell Change | Sheet1 | R3C3 | 1545 | <blank> | | |
| 5 | 2/13/2006 | 9:45 PM | Raymond Payette | Row Auto-Insert | Sheet1 | R4 | | | | |
| 6 | 2/13/2006 | 9:45 PM | Raymond Payette | Cell Change | Sheet1 | R4C3 | 1638 | <blank> | | |
| 7 | 2/13/2006 | 9:45 PM | Raymond Payette | Row Auto-Insert | Sheet1 | R5 | | | | |
| 8 | 2/13/2006 | 9:45 PM | Raymond Payette | Cell Change | Sheet1 | R5C3 | 1995 | <blank> | | |
| 9 | 2/13/2006 | 9:45 PM | Raymond Payette | Row Auto-Insert | Sheet1 | R6 | | | | |
| 10 | 2/13/2006 | 9:45 PM | Raymond Payette | Cell Change | Sheet1 | R6C3 | 2055 | <blank> | | |
| 11 | 2/13/2006 | 9:45 PM | Raymond Payette | Row Auto-Insert | Sheet1 | R7 | | | | |
| 12 | 2/13/2006 | 9:45 PM | Raymond Payette | Row Auto-Insert | Sheet1 | R8 | | | | |
| 13 | 2/13/2006 | 9:45 PM | Raymond Payette | Cell Change | Sheet1 | R8C3 | 1685 | <blank> | | |
| 14 | 2/13/2006 | 9:45 PM | Raymond Payette | Row Auto-Insert | Sheet1 | R9 | | | | |
| 15 | 2/13/2006 | 9:45 PM | Raymond Payette | Cell Change | Sheet1 | R9C3 | 1660 | <blank> | | |
| 16 | 2/13/2006 | 9:45 PM | Raymond Payette | Row Auto-Insert | Sheet1 | R10 | | | | |
| 17 | 2/13/2006 | 9:45 PM | Raymond Payette | Cell Change | Sheet1 | R10C3 | 1998 | <blank> | | |
| 18 | 2/13/2006 | 9:45 PM | Raymond Payette | Cell Change | Sheet1 | R12C3 | =SUM(R[-10]C:R[-1]C) | <blank> | | |
| 19 | 2/13/2006 | 9:45 PM | Raymond Payette | Cell Change | Sheet1 | R7C3 | $2,585.00 | <blank> | | |
| 20 | 2/13/2006 | 9:45 PM | Raymond Payette | Cell Change | Sheet1 | R2C3 | $1,385.00 | <blank> | | |

The history ends with the changes saved on 2/13/2006 at 9:45 PM.

Microsoft designed this tracking tool in the context of file sharing, not documenting. Macros cannot be viewed or modified when it is in sharing mode, so the tracking results cannot be modified for documenting purposes. As we can observe a single entry produces 11 possible entries and sometimes it doubles that amount. A 12 row common addition has become a 17 by 11 matrix (not counting the subsequent cell changes)! Obviously this analysis can become very tedious and understandably would require advanced databases to analyze and the Total Cost of Ownership would be very high. Perhaps there is a better way to do this!

A simple documentation would be to explain the purpose of what we did:

| | 1 | 2 | 3 |
|---|---|---|---|
| 1 | Items | Raymond Payette: Addition of data | |
| 2 | $ 1,385 | | |
| 3 | $ 1,545 | | |
| 4 | $ 1,638 | | |
| 5 | $ 1,995 | | |
| 6 | $ 2,055 | | |
| 7 | $ 2,585 | | |
| 8 | $ 1,695 | | |
| 9 | $ 1,660 | | |
| 10 | $ 1,998 | | |
| 11 | | | |
| 12 | $ 16,556 | | |
| 13 | | | |

We could complete the information, but one of the problems with documentation is that we find it tedious and we don't do this, so we could have a macro that does some documentation for us:

|   | 1 | 2 | 3 | 4 |
|---|---|---|---|---|
| 1 | Items | Author: Raymond Payette | | |
| 2 | $ 1,385 | Date: 02/16/2006 | | |
| 3 | $ 1,545 | Time: 9:45PM | | |
| 4 | $ 1,638 | | | |
| 5 | $ 1,995 | | | |
| 6 | $ 2,055 | | | |
| 7 | $ 2,585 | | | |
| 8 | $ 1,695 | | | |
| 9 | $ 1,660 | | | |
| 10 | $ 1,998 | | | |
| 11 | | | | |
| 12 | $ 16,556 | | | |
| 13 | | | | |

If we had a User Form with listboxes and textboxes they could easily make choices and it would make it easier to complete the documentation:

|   | 1 | 2 | 3 | 4 |
|---|---|---|---|---|
| 1 | Items | | | |
| 2 | 1,385.45 | Author: Raymond Payette | | |
| 3 | 1,545.65 | Date: 2/20/2006 | | |
| 4 | 1,325.30 | Time: 10:43 AM | | |
| 5 | 1,995.00 | Purpose: Daily cash receipts | | |
| 6 | 2,055.90 | Type: Data (Validated) | | |
| 7 | 2,585.25 | Source: Cashiers | | |
| 8 | 1,695.50 | Range: A2:A10 | | |
| 9 | 1,660.75 | Format: Currency | | |
| 10 | 1,998.10 | Checked by: Ben Jones | | |
| 11 | | Date: 2/21/2006 | | |
| 12 | 16,247 | Update: Daily | | |
| 13 | | Validation | | |
| 14 | | Amount between 0 and | | |
| 15 | | $10,000 | | |

This is much more significant and yet it is fairly easy to implement.

The above comments apply to data cells, but different types of cells could have different sorts of comments. There are 4 types of cells, **Titles**, **Data**, **Formulas** and **Links**. They have common information and they have specific information. For example all have an author and a date. A Title does not have a Source nor is it Updated.

Here is an example of Cell Documentation (except links that should be similarly documented):

[Screenshot of spreadsheet with cell comments showing Author: Raymond Payette, Date: 2/20/2006, Time: 10:43 AM, Purpose, Type, Source, Range, Format, Checked by, Date, Update fields for cells containing Items (1,385.45 etc.), totals (16,247), and validation between 0 and $10,000]

Though only 3 cells have been commented the comments contain much more useful information than other methods; moreover by standardizing the information, it facilitates the commenting process and it allows to produce standardized reports that can be quickly analyzed:

| | 1 | 2 | 3 | 4 | 5 | 6 | 7 | 8 | 9 | 10 | 11 |
|---|---|---|---|---|---|---|---|---|---|---|---|
| 1 | Author | Date | Time | Purpose | Type | Source | Range | Format | Checked by | Date | Update |
| 2 | Raymond Payette | 2/20/2006 | 10:43 AM | Add daily receipts | Title & Label | NA | NA | General | Ben Jones | 2/21/2006 | NA |
| 3 | Raymond Payette | 2/20/2006 | 10:43 AM | Daily cash receipts | Data (Validated) | Cashiers | A2:A10 | Currency | Ben Jones | 2/21/2006 | Daily |
| 4 | Raymond Payette | 2/20/2006 | 10:43 AM | Total od weekly receipts | Formula | NA | A2:A10 | Currency-roundedl | Ben Jones | 2/21/2006 | Weekly |

Data changes usually do not need to be commented. When Structural changes are made, they should be documented:

| | 1 | 2 | 3 | 4 | 5 | 6 | 7 | 8 | 9 | 10 | 11 |
|---|---|---|---|---|---|---|---|---|---|---|---|
| 1 | Author | Date | Time | Purpose | Type | Source | Range | Format | Checked by | Date | Update |
| 2 | Raymond Payette | 2/20/2006 | 10:43 AM | Add daily receipts | Title & Label | NA | NA | General | Ben Jones | 2/21/2006 | NA |
| 3 | Raymond Payette | 2/20/2006 | 10:43 AM | Daily cash receipts | Data (Validated) | Cashiers | A2:A10 | Currency | Ben Jones | 2/21/2006 | Daily |
| 4 | Raymond Payette | 2/20/2006 | 10:43 AM | Total of weekly receipts | Formula | NA | A2:A10 | Currency-rounded | Ben Jones | 2/21/2006 | Weekly |
| 5 | Raymond Payette | 2/26/2006 | 11:39 AM | Modification - not rounded. | Formula | NA | A2:A10 | Currency | | | Weekly |

Obviously this can be sorted and filtered by type, author or whatever is useful:

| | 1 | 2 | 3 | 4 | 5 | 6 | 7 | 8 | 9 | 10 | 11 |
|---|---|---|---|---|---|---|---|---|---|---|---|
| 1 | Author | Date | Time | Purpose | Type | Source | Range | Format | Checked by | Date | Update |
| 4 | Raymond Payette | 2/20/2006 | 10:43 AM | Total of weekly receipts | Formula | NA | A2:A10 | Currency-rounded | Ben Jones | 2/21/2006 | Weekly |
| 5 | Raymond Payette | 2/26/2006 | 11:39 AM | Modification - not rounded. | Formula | NA | A2:A10 | Currency | | | Weekly |

The different types of cells can be color-coded according to the specified Range. Using the Track Precedents, it can easily be shown that all the cells have been properly documented:

| | 1 |
|---|---|
| 1 | Items |
| 2 | 1,385.45 |
| 3 | 1,545.65 |
| 4 | 1,325.30 |
| 5 | 1,995.00 |
| 6 | 2,055.90 |
| 7 | 2,585.25 |
| 8 | 1,695.50 |
| 9 | 1,660.75 |
| 10 | 1,998.10 |
| 11 | |
| 12 | 16,247 |
| 13 | |

Data mapping can be useful to analyze specific aspects; a standardized information structure is essential. For example the Update frequency could be mapped:

| | 1 |
|---|---|
| 1 | NA |
| 2 | Daily |
| 3 | Daily |
| 4 | Daily |
| 5 | Daily |
| 6 | Daily |
| 7 | Daily |
| 8 | Daily |
| 9 | Daily |
| 10 | Daily |
| 11 | |
| 12 | Weekly |
| 13 | |

These possibilities considerably extend the usefulness of proper documentation.

**3.5 Documenting macros**

Excel's macros are one of its most powerful features. They can completely change the information and if incorrectly used they can become a source of undocumented errors. It is essential that they be fully understood and checked. To achieve this end, an information structure similar to the one devised for cells should be applied. For example:

|   | A | B | C | D |
|---|---|---|---|---|
| 1 | Amount |   |   |   |
| 2 | 1000 |   |   |   |
| 3 | 1250 |   |   |   |
| 4 | 1350 |   |   |   |
| 5 | 1250 |   |   |   |
| 6 | 1000 |   |   |   |
| 7 | 5750 |   |   |   |
| 8 |   |   |   |   |

A7  fx =SUM(A2:A6)

The total is incorrect, because cell A5 was changed from 1150 to 1250 and the following macro had previously been applied:

```
Private Sub Worksheet_Activate()
    With Application
        .Calculation = xlManual
        .MaxChange = 0.001
    End With
    ActiveWorkbook.PrecisionAsDisplayed = False
End Sub
```

The manual calculation will not update the total when a change is made!

In general macros should be systematically commented such as:

```
Private Sub Worksheet_Activate()
    'Author: Ken Jones
    'Date: 02/22/2006
    'Time: 9:38 AM
    'Purpose: To save the file every time the worksheet is used
    'Type: macro
    'Source: MyFile.xls, Sheet2, Worksheet_Activate
    'Range: Entire Workbook
    'Format: Excel file
    'Checked by: Raymond Payette
    'Date: 02/22/2006
    'Update: Review yearly
    ActiveWorkbook.SaveAs Filename:="C:\MyFile.xls"
End Sub
```

A separate sheet should analyze macros comments:

| | A | B | C | D | E | F | G | H | I | J | K |
|---|---|---|---|---|---|---|---|---|---|---|---|
| 1 | Author | Date | Time | Purpose | Type | Source | Range | Format | Checked by | Date | Update |
| 2 | Ken Jones | 2/22/2006 | 9:32 AM | To save the file every time the worksheet is used | Macro | MyFile.xls, Sheet2, Worksheet_Activate | Entire workbook | Excel file | Raymond Payette | 2/22/2006 | Review yearly |

Further comments should be added to explain the Sub-Procedure's parameters and logic.

## 4. CONCLUSION

Spreadsheet documentation doesn't have to be a daunting task to do or to analyze. Using Excel's comments in a systematic way can become an efficient documentation method. This in turn will help to do the change control by updating and analyzing the comments in a systematic way.

**6. Trade Marks:**

Excel®, SharePoint®, InfoPath®, Windows Server 2003® and SQL Server 2000®. Are trademarks of Microsoft Corporation.

SCANXLS add-in is a trademark of Systems Modeling Ltd  http://www.sysmod.com 02/20/2006  2:38pm

Blank page